\documentclass[12pt]{article}

\usepackage{natbib}
\usepackage{epsfig} 
\usepackage{amssymb}
\input{epsf}
\usepackage{graphicx}
\usepackage{amsthm}

\def\be{\begin{equation}}
\def\bea{\begin{eqnarray}}
\def\ee{\end{equation}}
\def\eea{\end{eqnarray}}

\usepackage{times}

\newenvironment{arxivabstract}{%
\begin{quote} \bf}
{\end{quote}}

\newcounter{lastnote}

\title{Schelling's Segregation Model: Parameters, Scaling, and Aggregation}

\author
{Abhinav Singh,$^{1}$ Dmitri Vainchtein,$^{1\ast}$ Howard Weiss$^{2}$\\
\\
\normalsize{$^{1}$School of Physics and Center for Nonlinear Science, Georgia Tech, USA}\\
\normalsize{$^{2}$School of Mathematics and Center for Nonlinear Science, Georgia Tech, USA}\\
\\
\normalsize{$^\ast$To whom correspondence should be addressed;
E-mail: dmitri@gatech.edu.}\thanks{This work was partially
supported by NSF (grants DMS-0355180 and 0400370) and by the Donors
of the ACS Petroleum Research Fund.}
}

\date{}

\begin{document}

\maketitle

\begin{arxivabstract}
Thomas Schelling proposed an influential simple spatial model to illustrate how,
even with relatively mild assumptions on each individual's nearest
neighbor preferences, an integrated city would likely unravel to a
segregated city, even if all individuals prefer integration.
Aggregation relates to individuals coming together to form groups and global aggregation corresponds to segregation. Many authors assumed that the segregation which Schelling observed in
simulations on very small cities persists for larger, realistic size
cities.  We devise new measures to quantify the
segregation and unlock its dependence on city size, disparate
neighbor comfortability threshold, and population density. We
identify distinct scales of global aggregation, and  show that the
striking global aggregation Schelling observed is strictly a small
city phenomenon. We also discover several  scaling laws
for the aggregation measures.

\end{arxivabstract}

\newpage

\section{Introduction}
In the 1970s, the eminent economic modeler Thomas Schelling proposed
a simple space-time population model to illustrate how, even with
relatively mild assumptions concerning every individual's nearest
neighbor preferences, an integrated city would likely unravel to a
segregated city, even if all individuals prefer integration
\cite{S1,S2,S3,S4}. His agent based lattice model has become
quite influential amongst social scientists, demographers, and
economists, and a number of authors are testing the Schelling model
using actual population data \cite{Clark,BR,BOHO,SSD}. The only
quantitative analysis of such models we could locate in the
literature are \cite{PW} and \cite{GGSWZ}.

Aggregation relates to individuals coming together to form groups or
clusters according to race, and global aggregation corresponds to
segregation. Many authors assume that the striking global
aggregation observed in simulations on very small ideal ``cities"
persists for large, realistic size cities. A recent paper \cite{VK}
exhibits final states for a small number of model simulations of a
large city, and some final states that do not exhibit significant
global aggregation. However, quantification of this important
phenomenon is lacking in the literature, presumably, in part, due to
the large computational costs required to run simulations using
existing algorithms.

In this paper, we devise new measures to quantify the aggregation
and unlock its dependence on city size, disparate neighbor
comfortability threshold, and population density. We developed a
highly efficient and fast algorithm that allows us to compute
meaningful statistics of these measures(see Figures~\ref{clust34},\ref{unhappy},\ref{Jumpsstat}). We identify distinct scales
of global aggregation, and we show that the striking global
aggregation Schelling observed is strictly a small city phenomenon.
We also discover several remarkable scaling laws for the aggregation
measures.

\subsection{Description of the Model}

We expand Schelling's original model \footnote{Different authors
frequently consider slightly different versions of Schelling's
original model, i.e., different ways of moving boundary agents. All
versions seem to exhibit the same qualitative behaviors, and thus we
refer to {\it the} Schelling model.} to a three parameter family of
models. The phase space for these models is the $N \times N$ square
lattice with periodic boundary conditions. We consider two distinct
populations, that, in Schelling's words \cite{S4}, refer to
``membership in one of two homogeneous groups: men or women, blacks
and whites, French-speaking and English speaking, officers and
enlisted men, students and faculty, surfers and swimmers, the well
dressed and the poorly dressed, and any other dichotomy that is
exhaustive and recognizable.'' We denote by B (black squares) and R
(red squares) these two populations. See Figure 1. Together these
agents fill up some of the $N^2$ sites, with $V$ remaining vacant
sites (white squares). Each agent has eight nearest neighbors (Moore
neighborhood). Fix a {\it disparate neighbor comfortability
threshold} $T \in \{0, 1, \dots, 8\}$, and declare that a B or R is
{\it happy} if $T$ or more of its nearest eight neighbors are B's or
R's, respectively. Else it is unhappy.

Demographically, the parameter $N$ controls the size of the city,
$v=V/N^2$ controls the population density or the {\it occupancy
ratio} \cite{realestate}, and $T$ is an ``agent comfortability
index'' that quantifies an agent's tolerance to living amongst
disparate nearest neighbors.

We begin the evolution by choosing an initial configuration
(described in Sect.~3) and randomly selecting an unhappy B and a
vacant site surrounded by at least $T$ nearest B neighbors. Provided
this is possible, interchange the unhappy B with the vacant site, so
that this B becomes happy. Then randomly select an unhappy R and a
vacant site having at least $T$ nearest neighbors of
 type R. Provided this is possible, interchange the unhappy R
with the vacant site, so that R becomes happy. Repeat this iterative
procedure, alternating between selecting an unhappy B and an unhappy
R, until a {\it final state} is reached, where no interchange is
possible that increases happiness. For some final states, some (and
in some cases, many) agents may be unhappy, but there are no
allowable switches.

\section{Schelling's results}

\begin{figure}[htbp]
\center\begin{tabular}{cc}
\vspace*{0mm}\includegraphics[height=2in]{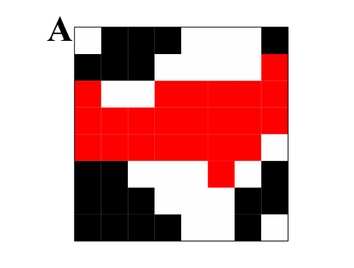}&
\hspace*{10mm}\includegraphics[height=2in]{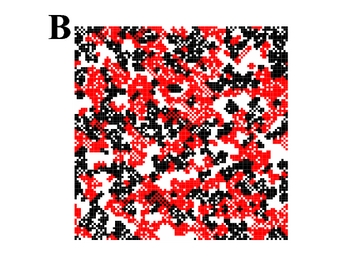}
\end{tabular}
\vspace*{0mm} \caption{\label{f1} {\bf A}: A simulation of
Schelling's original model with $N=8$; {\bf B}: Our simulation with
$N=100$.}
\end{figure}

Schelling considered the case $N=8, T=3,$ and $v = 33\%$. The
details of Schelling's simulations (including the algorithm for
choosing initial conditions) are presented in Sec.~\ref{Schell} in
the supporting materials. The final state of a typical run of
Schelling's original model system is presented in Figs.~\ref{f1}.
Schelling performed many simulations by hand using an actual
checkerboard, and observed that the final states presented a
significant degree of global aggregation. He equated the global
aggregation with segregation of a city.

In this paper, we investigate whether the global aggregation that
Schelling observed for very small lattices persists for larger
lattices. In Fig.~\ref{f1}{\bf B} , we present a characteristic
final state for our simulations with $N=100$. Comparing
Figs.~\ref{f1}{\bf A} and ~\ref{f1}{\bf B} , one can see a striking
qualitative difference between the two final states. While there is
some local aggregation in the final state with $N=100$, there is no
global aggregation. By viewing the plots of this and other final
states, one immediately sees that the global aggregation that
Schelling observed is a small lattice phenomenon. In Sections 3
through 5 we quantify the global aggregation using several new
measures, and we carefully analyze the structure of the final states
for different values of $T$ and $v$. In particular, we quantify the
claim that global aggregation is a small lattice phenomenon.

\section{Simulations}

We study the dynamics for large lattices and present our results for
$N=100$. Figures 2-10 (except Fig.~\ref{fsm1}) are all based on
$N=100$. In the supporting materials, we discuss the cases $N=50$
and $N=200$, and claim that $N$ greater than $100$ does not lead to
qualitatively or quantitatively different states and phenomena. We
restrict our discussion to cities having an equal number of B's and
R's. A separate manuscript \cite{SVW} will study the dynamics with
different proportions of B's and R's.

We consider $T=3, 4, 5$ and $v$ between $2\%$ and $33\%$. The system
does not evolve very much for other values of $T$: for $T=1, 2$
almost all of the agents are satisfied in most of the initial
configurations, while for $T \ge 6$ there are almost no legal
switches. Values of $v$ larger then $33\%$ correspond to unrealistic
environments. For each pair of parameters $T$ and $v$, we perform
$100$ simulations (statistics are presented below and in the
supporting materials). This number of simulations was chosen to
ensure a 95\% confidence interval for parameter estimation.

We choose the initial configuration by starting with a checkerboard
with periodic boundary conditions. Demographically, a checkerboard
configuration is a maximally integrated configuration. We then
randomly remove $V/2$ of both B's and R's (thus keeping equal
numbers of both agents). We permute agents in two $3 \times 3$
blocks. Alternatively, we could choose a random initial
configuration. In general, except for small $v$, the final states
are quantitatively similar as for our Schelling initial conditions.

\begin{figure}[htbp]
\center\begin{tabular}{ccc}
\hspace*{-18mm}\includegraphics[width=2.8in]{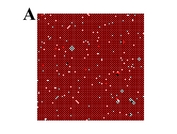}&
\hspace*{-18mm}\includegraphics[width=2.8in]{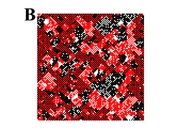}&
\hspace*{-18mm}\includegraphics[width=2.8in]{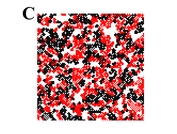} \\
\hspace*{-18mm}\includegraphics[width=2.8in]{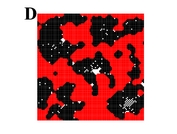} &
\hspace*{-18mm}\includegraphics[width=2.8in]{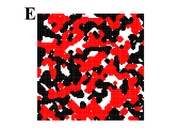} &
\hspace*{-18mm}\includegraphics[width=2.8in]{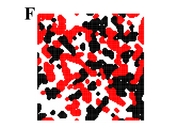} \\
\hspace*{-18mm}\includegraphics[width=2.8in]{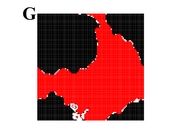} &
\hspace*{-18mm}\includegraphics[width=2.8in]{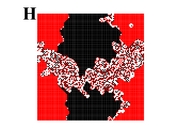} &
\hspace*{-18mm}\includegraphics[width=2.8in]{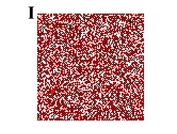}
\end{tabular}
\caption{\label{fN2} Characteristic final states for different
values of $T$ and $v$: {\bf A}: $T=3$, $v=2\%$, {\bf B}: $T=3$,
$v=15\%$, {\bf C}: $T=3$, $v=33\%$, {\bf D}: $T=4$, $v=2\%$, {\bf
E}: $T=4$, $v=15\%$, {\bf F}: $T=4$, $v=33\%$, {\bf G}: $T=5$,
$v=2\%$, {\bf H}: $T=5$, $v=15\%$, {\bf I}: $T=5$, $v=33\%$.}
\end{figure}

In Fig.~\ref{fN2}, we present characteristic final states for
different values of $T$ and $v$. Visually, the aggregation in the
final states along each column (with fixed $v$) is substantially
different than in the final states along each row (with fixed $T$).
More characteristic final states are presented in
Figs.~\ref{T3varV}-\ref{T5varV} in the supporting materials . We
begin the next section by defining several measures of aggregation
that enable us to quantify this observation.

\section{Analysis}

\subsection{Measures of Aggregation}

One can see from Fig.~\ref{fN2} that aggregation is a multifaceted
phenomenon, and thus requires several measures to describe.
Schelling used two measures to quantify the aggregation: (1) the
average over all agents of the quotient of the number of like to
unlike neighbors, and (2) the number of agents whose eight nearest
neighbors all have the same label, and thus are completely
surrounded by like agents. We call the latter quantity {\it
seclusiveness} and denote by $N_0$.

We introduce three additional measures. Aggregation manifests itself
in two ways: (i) reduced number of contacts between agents of
different kinds and (ii) the dominance of agents of either kind in a
certain area. The first of the following measures of aggregation
quantify (i) and the second and third measures quantify (ii).

\medskip

\noindent (3) {\it The adjusted perimeter per agent $p$ of the
interface between the different agents suitably adjusted for the
vacant spaces.} The adjusted perimeter is defined as twice the total
number of R-B connections plus the total number of connections
between R and B agents with vacant spaces, all divided by the total
number of agents $N^2$ (see the supporting materials).
Demographically, $p$ is the average number of contacts an agent has
with the opposite kind or with vacant sites.

\medskip

\noindent (4) {\it The scale, or maximum diameter of the connected
components (which we henceforth call clusters) $L$.} The measure $L$
is defined as the side length of the smallest square needed to cover
every cluster. For states consisting of compact clusters, larger
values of $L$ correspond to larger {\it scales of aggregation}.
Checkerboard configurations and configurations consisting of compact
clusters are two extremes; for the former $L=N$.

\medskip

\noindent (5) {\it The total number of clusters in a configuration
$N_C$.} This intuitively appealing measure of aggregation is useful
to describe final states having mostly large compact clusters. To
see its limitation, observe that ``the maximally integrated''
checkerboard configuration with $v=0$ has just two clusters.

\medskip

The definition of $p$ was motivated by analogies of these models
with the physics of foams. A key observation is that $p$ is a
Lyapunov function, i.e., a function defined on every configuration,
and which is strictly decreasing along the evolution of the system
until it reaches a final state (see proof in the supporting materials). Thus the system evolves to minimize the adjusted
perimeter between the interface of the R and B agents. The final
states are precisely the local minimizers of the Lyapunov function,
subject to the threshold constraint. This Lyapunov function is also
the Hamiltonian for a related spin lattice system related to the
Ising model \cite{Simon}.

\begin{figure}[htbp]
\center\begin{tabular}{cc}
\hspace*{-0mm}\includegraphics[width=3in]{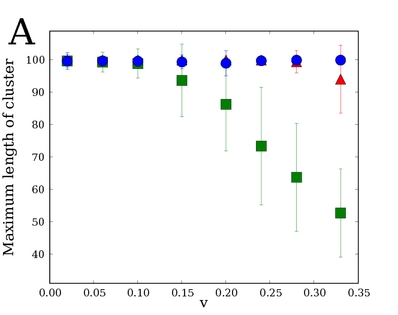}&
\hspace*{-0mm}\includegraphics[width=3in]{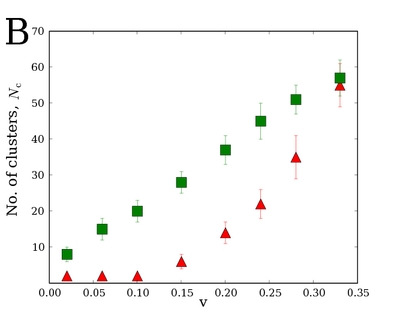} \\
\hspace*{0mm}{\bf } &\hspace*{-5mm}{\bf } \\
\hspace*{-0mm}\includegraphics[width=3in]{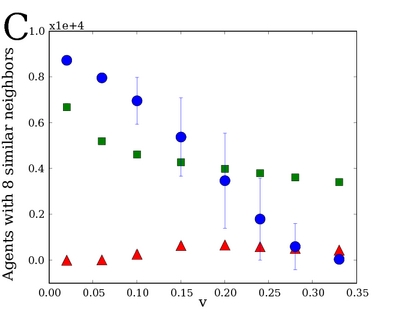} &
\hspace*{-0mm}\includegraphics[width=3in]{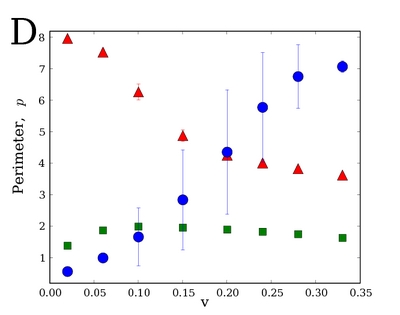} \\
\hspace*{0mm}{\bf } &\hspace*{-5mm}{\bf }
\end{tabular}
\vspace*{0mm} \caption{\label{clust34} Statistics of four key
measures of aggregation of final states for $T=3$ (red triangles),
$T=4$ (green squares), and $T=5$ (blue circles) for different $v$:
{\bf A} The scale of aggregation $L$; {\bf B} The number of clusters
$N_C$; {\bf C} The number of agents with eight like nearest
neighbors $N_0$; {\bf D} Normalized perimeter $p$.}
\end{figure}

In Fig.~\ref{clust34}, we plot average values of aggregation measures
(2)-(5) introduced above for the final states with $T=3,4,5$ and
several values of $v$. The linear relationships of these
disparate aggregation measures on population density seems remarkable.

\subsection{Global Aggregation Dependence on $\mathbf{T}$}

From Fig.~\ref{fN2}, one observes that: (i) the final states with
$T=3$ are very sparse, with a great deal of interweaving between
both kinds of agents and vacant spots; (ii) the final states with
$T=4$ consist of compact clusters (that look like ``solid''
objects); and, finally, (iii) the final states with $T=5$ consist of
one (for each type) huge cluster together with a small number of
remaining agents scattered around. Varying the density $v$ does not
radically alter the qualitative structure of the final states. We
now quantify the aggregation for each value of $T$, as $v$ varies
between 2\% and 33\%.

\subsubsection{$\mathbf{T=3}$: sparse clusters}

The sparsity of the final states with $T=3$ is due, in part, to
large blocks of the initial checkerboard configuration that remain
unchanged during the evolution. We call this phenomenon the {\it
super-stability} of the checkerboard. Every agent is not just happy,
but has one like neighbor to ``spare''. Thus, it takes a large
perturbation to make a given agent move and, therefore, only agents
close to the initially perturbed sites move. Consequently, Schelling
required a large density of vacant spaces $v$ ($33\%$) to overcome
checkerboard super-stability. One can see from the left column in
Fig.~\ref{fN2} and from Fig.~\ref{T3varV} in the supporting materials,
that as $v$ decreases, larger and larger parts of the initial
configuration remain unchanged during the evolution.

We observe that for $T=3$, larger values of $v$ result in larger
clusters, and thus lead to greater global aggregation. The number of
clusters in the final states, $N_C$, decreases as $v$ decreases
(Fig.~\ref{clust34}{\bf B}) and the dependence is almost cubic (see
Fig.~\ref{Tlog} in the supporting materials). The seclusiveness
measure, $N_0$, is a monotonically decreasing function of $v$: as
$v$ decreases, the final state approaches the checkerboard and,
naturally, almost all the agents have some contacts with other
agents. Similarly, the smaller the value of $v$, the larger the
normalized perimeter, $p$.

\subsubsection{$\mathbf{T=4}$: compact clusters and mesoscale aggregation}

Increasing $T$ from $3$ to $4$ eliminates the checkerboard
super-stability phenomenon and results in strikingly different
structures of aggregation in final states. Namely, the final states
consist of relatively small numbers, that clearly depend on $v$, of
compact clusters (see Fig.~\ref{fN2}{\bf D} -{\bf F} and
Fig.~\ref{T4varV} in the supporting materials).

For relatively large $v$, such final states exhibit {\it mesoscale
aggregation} and, for small values of $v$, {\it macroscale
aggregation}. There seems to be no canonical way to separate the two
types of aggregation. Our criterion to define the transition when
the size of the largest cluster, $L$, becomes equal to $N$.

We find two measures that clearly differentiate the global
clustering of the final states for $T=3$ and $T=4$. First, the final
states have very different perimeters (see Fig.~\ref{clust34}{\bf D}
). Second, for $T=3$, the clusters are very sparse, while for $T=4$,
the clusters are compact. A natural way to quantify this is to use
$N_0$, whose statistics we present in Fig.~\ref{clust34}{\bf C} .
The measure $N_0$ is a monotonically increasing function of $v$.

We quantify the differences in the final states for $T=4$ with $v$
ranging from $v=33\%$ down to $v=2\%$, with three measures: the
number of clusters, $N_C$, the scale of aggregation, $L$, and the
seclusiveness, $N_0$ (Fig.~\ref{clust34}). By providing
opportunities for increasingly ``easier satisfaction," one might
believe that decreasing $v$ increases the values of $N_C$. Our study
confirms this, and the dependence is remarkably linear. Namely, for
 typical final states with $T=4, v=33\%$ (Fig.~\ref{fN2}{\bf F}) $N_C$
is relatively high; for $T=4, v=15\%$, $N_C$ is smaller
(Fig.~\ref{fN2}{\bf E}) and the clusters on average are bigger;
finally, states with $T=4, v=2\%$ contain only a few compact
clusters of either type that stretch across the whole lattice
($L=100$). In general, as $v$ decreases, $L$ increases almost
linearly (see Fig.~\ref{clust34}{\bf A}).

{\it Thus for $T=3$ and $T=4$, the increase in $v$ leads to the
opposite effects: they {\it increase} and {\it decrease} the level of
global aggregation, respectively.}

\subsubsection{$\mathbf{T=5}$: final states with many unhappy agents}

For small $v$, the dynamics with $T=5$ results in a final state
achieved after just a few switches, and consists of mostly unhappy
agents with no vacant space to where they could move. However, a
slight modification of the selection algorithm, to allow direct R-B
switches when it is not possible to switch with a vacant space,
results in significant global aggregation and drastically reduces
the number of unhappy agents, although not eliminating them
entirely. The presence of unhappy agents in the final states is a
new phenomenon, which we do not observe in simulations for $T=3$
(while such configurations theoretically exist, they are extremely
unlikely) and is much less pronounced for $T=4$ (see
Fig.~\ref{unhappy}{\bf A} in the supporting materials).

A typical $T=5$ final state with modified selection algorithm
consists of one big cluster for each kind of agent and the rest of
the agents are unhappy and scattered around. The smaller that the
value of $v$ is, the larger the two main clusters are and the fewer
unhappy agents there are.

The globally aggregated final states (small values of $v$) with
$T=5$ (with modified selection) and $T=4$ (with Schelling selection)
appear similar in terms of the number of large clusters and the
scale of aggregation, $L$ (Figs.~\ref{fN2}{\bf D} and {\bf G}).
However, there is a large difference in their perimeter $p$: it is
much smaller for $T=5$ (see Fig.~\ref{clust34}{\bf D}).

\medskip

\section{Concluding Remarks}
In this paper we devised new measures to quantify the aggregation in
the Schelling segregation model and studied their dependence on city
size, disparate neighbor comfortability threshold, and population
density. We identified distinct scales of global aggregation, and
showed that the striking global aggregation Schelling observed for
the $8 \times 8$ lattice is strictly a small lattice phenomenon. We
also discovered several remarkable scaling laws for the aggregation
measures as functions of population density.

\medskip

\section{Supporting  Material}

\subsection{Schelling's Simulations}{\label{Schell}}

Schelling considered the cases $N=8, T=3$, and $v = 33\%$. For
$T=3$ or $4$, and $V=0$, a ``checkerboard" configuration of B's and
R's (imagine placing B's on the red squares and R's on the black
squares of an actual checkerboard) is a final state, since all
agents have four like nearest neighbors.

To generate his initial configurations, Schelling begins with a
checkerboard configuration without periodic boundary conditions and
randomly removes approximately one third of the B's and R's, keeping
equal numbers of both agents \cite{S4}. We refer to the result as a
deleted checkerboard configuration. Removing these agents makes some
of the remaining agents unhappy and drives the evolution. Several
authors have observed that removing such a large percentage of
agents is unnatural, but it is crucial to attain aggregation in
Schelling's model. Removing fewer agents results is a final
configuration close to the initial configuration.

Finally, to construct his initial configurations, Schelling modifies
the deleted checkerboard by randomly adding a total of $5$ B's and
$5$ R's in vacant spaces. Such an initial configuration is assumed
to be a proxy for a nearly integrated city.

\begin{figure}[htbp]
\center\begin{tabular}{cc}
\vspace*{0mm}\includegraphics[width=2in]{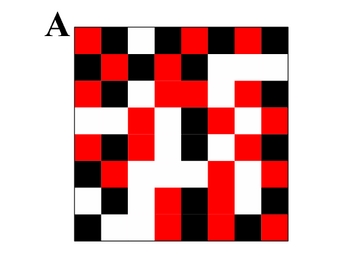}&
\hspace*{10mm}\includegraphics[width=2in]{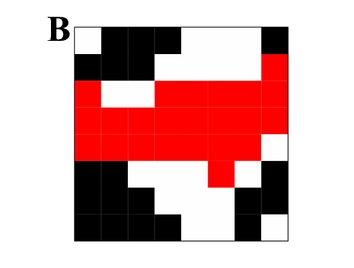}
\end{tabular}
\vspace*{0mm} \caption{\label{fsm1} Schelling's original model with
$N=8$: {\bf A}: initial state and {\bf B}: final state.}
\end{figure}

A typical run of Schelling's original model system (with periodic
boundary conditions) is presented in Fig.~\ref{fsm1}: initial state
(Fig.~\ref{fsm1}{\bf A}) and final state (Fig.~\ref{fsm1}{\bf B}).
Schelling performed many simulations by hand using an actual
checkerboard, and observed that the final states presented a
significant degree of global aggregation. He equated the global
aggregation with segregation of a city.

\subsection{Structure of final states for $\mathbf{T=3}$}

\begin{figure}[htbp]
\center\begin{tabular}{ccc}
\hspace*{-18mm}\includegraphics[width=2.8in]{ft3/final02N1.jpg}&
\hspace*{-18mm}\includegraphics[width=2.8in]{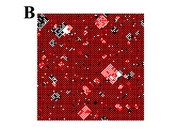} &
\hspace*{-18mm}\includegraphics[width=2.8in]{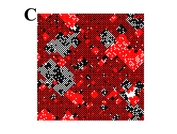} \\
\hspace*{-18mm}\includegraphics[width=2.8in]{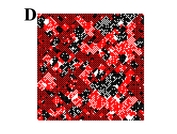}&
\hspace*{-18mm}\includegraphics[width=2.8in]{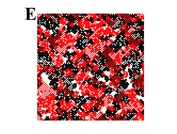} &
\hspace*{-18mm}\includegraphics[width=2.8in]{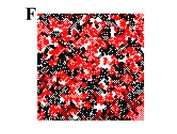} \\
\hspace*{-18mm}\includegraphics[width=2.8in]{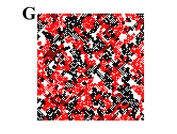}&
\hspace*{-18mm}\includegraphics[width=2.8in]{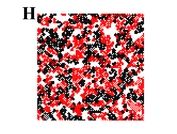} &
\end{tabular}
\caption{\label{T3varV} Characteristic final states for $T=3$ for
different $v$: {\bf A}: $v=2\%$, {\bf B}: $v=6\%$, {\bf C}:
$v=10\%$, {\bf D}: $v=15\%$, {\bf E}: $v=20\%$, {\bf F}: $v=24\%$,
{\bf G}: $v=28\%$, {\bf H}: $v=33\%$.}
\end{figure}

\begin{figure}[htbp]
\center\begin{tabular}{ccc}
\hspace*{-18mm}\includegraphics[width=2.8in]{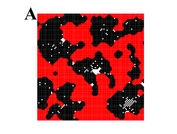}&
\hspace*{-18mm}\includegraphics[width=2.8in]{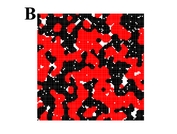} &
\hspace*{-18mm}\includegraphics[width=2.8in]{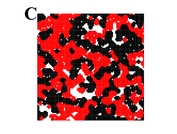} \\
\hspace*{-18mm}\includegraphics[width=2.8in]{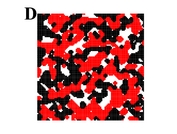}&
\hspace*{-18mm}\includegraphics[width=2.8in]{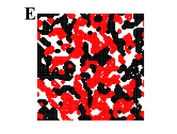} &
\hspace*{-18mm}\includegraphics[width=2.8in]{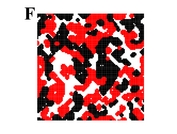} \\
\hspace*{-18mm}\includegraphics[width=2.8in]{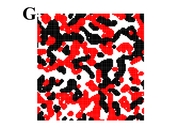}&
\hspace*{-18mm}\includegraphics[width=2.8in]{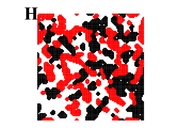} &
\end{tabular}
\caption{\label{T4varV} Characteristic final states for $T=4$ for
different $v$: {\bf A}: $v=2\%$, {\bf B}: $v=6\%$, {\bf C}:
$v=10\%$, {\bf D}: $v=15\%$, {\bf E}: $v=20\%$, {\bf F}: $v=24\%$,
{\bf G}: $v=28\%$, {\bf H}: $v=33\%$.}
\end{figure}

In Fig.~\ref{T3varV} we present typical final states for $T=3$ and
different values of $v$. The pictures and our statistics show that
as the value of $v$ decreases, the number of clusters decreases
(Fig.~\ref{clust34}{\bf B}), the final states retain larger and
larger parts of the original checkerboard configuration. One can see
from Fig.~\ref{T3varV} that, unlike the final states for Schelling's
original model with $N=8$ (see Fig.~\ref{fsm1}{\bf B}) that consists
of just one or two separate domains of R and B agents, the final
states for $N=100$ and $v=33\%$ contain many clusters. The striking
qualitative difference is also quantified by the relatively large
values of the normalized perimeter, $p$ (see Fig.~\ref{clust34}{\bf
B}) and by the large differences in the values of $N_0/N^2$ between
the states presented in Fig.~\ref{clust34}{\bf B} and
~\ref{fsm1}{\bf B}.

\subsection{Structure of final states for $\mathbf{T=4}$}

In Fig.~6 we present typical final states for $T=4$ and different
values of $v$. The pictures and our statistics show that as the
value of $v$ decreases, the number of clusters decreases
(Fig.~\ref{clust34}{\bf B}) and the size of the compact clusters
(scale of aggregation) increases (Fig.~\ref{clust34}{\bf A}).

\subsection{Structure of final states for $\mathbf{T=5}$}

\begin{figure}[htbp]
\center\begin{tabular}{ccc}
\hspace*{-18mm}\includegraphics[width=2.8in]{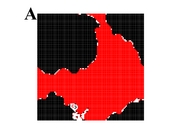}&
\hspace*{-18mm}\includegraphics[width=2.8in]{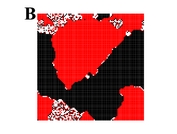} &
\hspace*{-18mm}\includegraphics[width=2.8in]{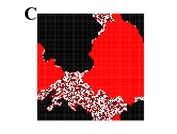} \\
\hspace*{-18mm}\includegraphics[width=2.8in]{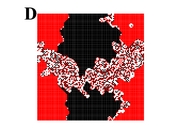}&
\hspace*{-18mm}\includegraphics[width=2.8in]{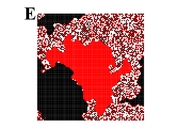} &
\hspace*{-18mm}\includegraphics[width=2.8in]{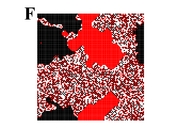} \\
\hspace*{-18mm}\includegraphics[width=2.8in]{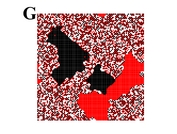}&
\hspace*{-18mm}\includegraphics[width=2.8in]{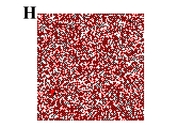} &
\end{tabular}
\caption{\label{T5varV} Characteristic final states for $T=5$ for
different $v$: {\bf A}: $v=2\%$, {\bf B}: $v=6\%$, {\bf C}:
$v=10\%$, {\bf D}: $v=15\%$, {\bf E}: $v=20\%$, {\bf F}: $v=24\%$,
{\bf G}: $v=28\%$, {\bf H}: $v=33\%$.}
\end{figure}

\begin{figure}[t]
\center\begin{tabular}{cc}
\hspace*{-0mm}\includegraphics[width=2in]{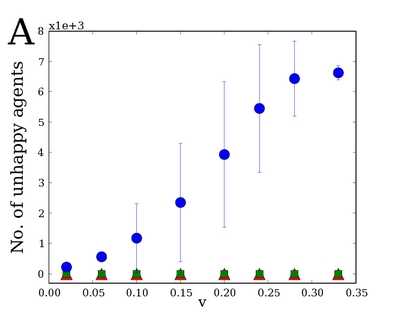}&
\hspace*{-0mm}\includegraphics[width=2in]{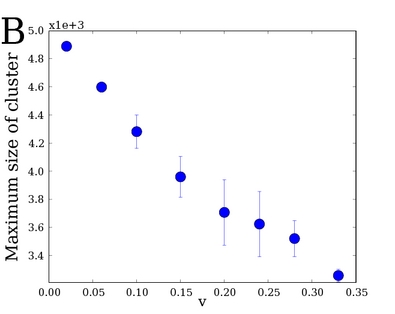}
\end{tabular}
\vspace*{0mm} \caption{\label{unhappy} Statistics of the final
states with $T=5$: {\bf A}: The average number of unhappy agents in
final states for $T=3$ (red triangles), $T=4$ (green squares), and
$T=5$ (blue circles); {\bf B}: The average number of the agents in
the two big clusters.}
\end{figure}

In Fig.~\ref{T5varV} we present typical final states for $T=5$ and
different values of $v$. Observe that most of the agents away from
the big clusters are unhappy (see Fig.~\ref{unhappy} for the
statistics of the unhappy agents). The pictures and our statistics
show that as the value of $v$ decreases, the number of unhappy
agents in final states $v$ decreases (Fig.~\ref{unhappy}{\bf A}),
and the size of a single (for each type) major cluster increases
(see Fig.~\ref{unhappy}{\bf B}). Another clear indication of the
growth of the main cluster is the increase of $N_0$, presented in
Fig.~\ref{clust34}{\bf C}.

The average number of unhappy agents in final states for different
values of $T$ and $v$ is presented in Fig.~\ref{unhappy}{\bf A} . It
is remarkable that the average number of unhappy agents is almost a
linear function of $v$, between $v=10\%$ (where they constitute
approximately $10\%$) and $v=30\%$ (where they constitute
approximately $33\%$, in other words, almost every agent). While the
existence of unhappy agents in the final state does not
significantly increase the perimeter $p$ of the final states, it
greatly inflates the total number of clusters $N_C$.

\subsection{Statistics of the number of clusters}

\begin{figure}[t]
\center\includegraphics[height=2in]{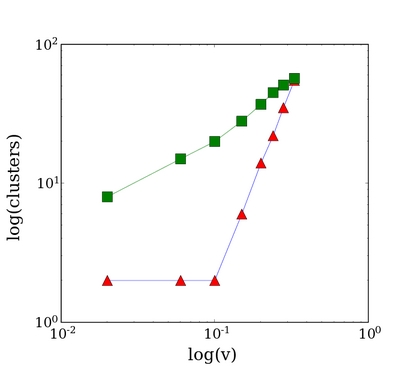}
\caption{\label{Tlog} Statistics of the number of clusters in the
final states for $T=3$ (red triangles), $T=4$ (green squares) for
different $v$ on a $\log-\log$ scale.}
\end{figure}

In Fig.~\ref{Tlog} we replot the data shown in
Fig.~\ref{clust34}{\bf B} on a log scale (to illustrate power-law
statistics), showing the statistics of the number of clusters in the
final states for $T=3$ and $T=4$, for different $v$. The lines
indicate power laws. The values for the slopes are $0.89$ for $T=4$
(the value $1$ is well within the error bars) and $2.76$ for $T=3$
(the value $3$ is well within the error bars).

\subsection{Number of steps in the evolution}

\begin{figure}[t]
\center\includegraphics[height=3in]{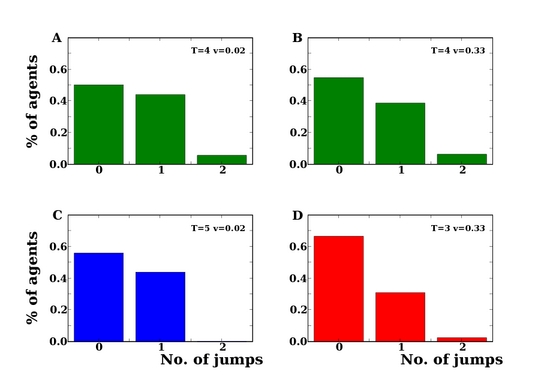}
\caption{\label{Jumpsstat} Statistics of the number of agent jumps.}
\end{figure}

The average number of switches required for the final state to be
achieved are (3596, 5192, 5573, 4422) respectively, for $T=3,
v=33\%$; $T=4, v=33\%$; $T=4, v=2\%$; and $T=5, v= 2\%$. The
switches for $T=5, v= 0.02$ that included both R and B agents are
counted twice. The most striking feature is that it takes
significantly fewer switches to achieve the final state for $T=5$
than for $T=4$. The distribution of the number of jumps for
different agents is presented in Fig.~\ref{Jumpsstat}.

\subsection{Final states with $\mathbf{N=50}$ and $\mathbf{N=200}$}

\begin{figure}[htbp]
\center\begin{tabular}{ccc}
\hspace*{-18mm}\includegraphics[width=2.8in]{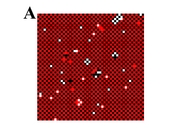}&
\hspace*{-18mm}\includegraphics[width=2.8in]{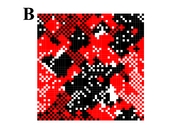}&
\hspace*{-18mm}\includegraphics[width=2.8in]{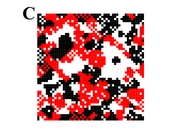} \\
\hspace*{-18mm}\includegraphics[width=2.8in]{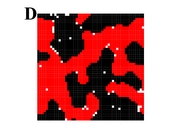} &
\hspace*{-18mm}\includegraphics[width=2.8in]{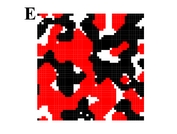} &
\hspace*{-18mm}\includegraphics[width=2.8in]{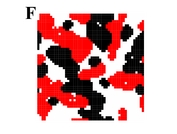} \\
\hspace*{-18mm}\includegraphics[width=2.8in]{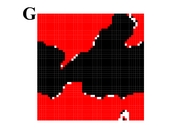} &
\hspace*{-18mm}\includegraphics[width=2.8in]{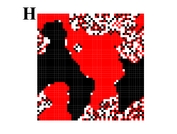} &
\hspace*{-18mm}\includegraphics[width=2.8in]{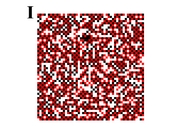}
\end{tabular}
\caption{\label{fN50} Characteristic final states for $N=50$ and
different values of $T$ and $v$: {\bf A}: $T=3$, $v=2\%$, {\bf B}:
$T=3$, $v=15\%$, {\bf C}: $T=3$, $v=33\%$, {\bf D}: $T=4$, $v=2\%$,
{\bf E}: $T=4$, $v=15\%$, {\bf F}: $T=4$, $v=33\%$, {\bf G}: $T=5$,
$v=2\%$, {\bf H}: $T=5$, $v=15\%$, {\bf I}: $T=5$, $v=33\%$.}
\end{figure}

\begin{figure}[htbp]
\center\begin{tabular}{ccc}
\hspace*{-18mm}\includegraphics[width=2.8in]{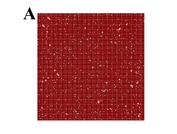}&
\hspace*{-18mm}\includegraphics[width=2.8in]{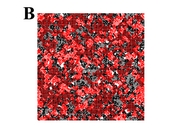}&
\hspace*{-18mm}\includegraphics[width=2.8in]{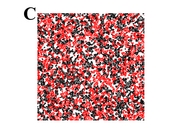} \\
\hspace*{-18mm}\includegraphics[width=2.8in]{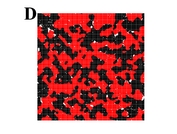} &
\hspace*{-18mm}\includegraphics[width=2.8in]{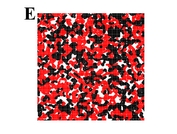} &
\hspace*{-18mm}\includegraphics[width=2.8in]{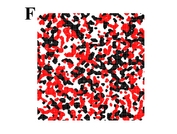} \\
\hspace*{-18mm}\includegraphics[width=2.8in]{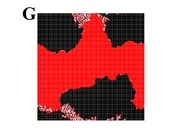} &
\hspace*{-18mm}\includegraphics[width=2.8in]{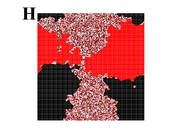} &
\hspace*{-18mm}\includegraphics[width=2.8in]{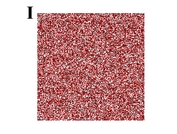}
\end{tabular}
\caption{\label{fN200} Characteristic final states for $N=200$ and
different values of $T$ and $v$: {\bf A}: $T=3$, $v=2\%$, {\bf B}:
$T=3$, $v=15\%$, {\bf C}: $T=3$, $v=33\%$, {\bf D}: $T=4$, $v=2\%$,
{\bf E}: $T=4$, $v=15\%$, {\bf F}: $T=4$, $v=33\%$, {\bf G}: $T=5$,
$v=2\%$, {\bf H}: $T=5$, $v=15\%$, {\bf I}: $T=5$, $v=33\%$.}
\end{figure}

To illustrate the dependence of the final states of $N$, we
performed the simulations for $N=50$ and for $N=200$. Characteristic
final states are presented in Fig.~\ref{fN50} and Fig.~\ref{fN200}
for $N=50$ and $N=200$, respectively. One can see that both figures
are qualitatively very similar to Fig.~\ref{fN2}.

\subsection{Measures of Aggregation}

Schelling considered two quantities to measure the aggregation of a
state:

\noindent 1. The ratio of unlike to like neighbors is called the
$[u/l]$-measure. For a site on the final lattice with coordinates
$(i, j)$), we define:
\[
[u/l]_{i,j} = \frac{q_{i,j}+w_{i,j}}{p_{i,j}},
\]
where $p_{i,j}$, $q_{i,j}$, and $w_{i,j}$ are the number of like,
unlike, and vacant neighbors of the agent located at $(i,j)$,
respectively. We define the {\it sparsity} $\left<[u/l]\right>$ of a
cluster by averaging the $[u/l]$-measure over the given cluster.

\noindent 2. The number of agents that have neighbors {\it only} of
the same kind, $N_0$ (note, that this definition excludes the vacant
spaces as well). The abundance of such agents indicate the presence
of large, ``solid'' clusters. This quantity is the most useful in
quantifying between the states with $T=3$ and $T=4$.

\medskip

\noindent Besides those two characteristics, we introduce several
more.

\medskip

\noindent 3. The total number of clusters, $N_C$. With 8-point
neighborhoods, clusters may be intermingled. For example, a
checkerboard configuration has just $1+1=2$ clusters. The quantity
$N_C$ is the most useful for configurations consisting of compact
clusters of a similar size. To study configurations such as the
final states for $T=5$, a more useful quantity is the number of
clusters that have greater then, say, $M_{max}/10$ agents, where
$M_{max}$ is the number of agents in the largest cluster of a given
kind.

\medskip

\noindent 4. Total perimeter (normalized by the number of agents).
For $v\ne 0$, we follow the calculation of surface tension in the
physics of foams, and write
\[
p = \frac{P}{N^2} = \frac1{N^2} \sum_{i,j=1}^{N} \left( q_{i,j} +
w_{i,j}/2 \right).
\]
Thus defined, $P$ plays the role of a {\it Lyapunov function}: it
can be shown that in the process of evolution every legal switch
makes $P$ smaller.

Suppose we switch an R and a V. Before the switch, suppose R had
$B_1$, $R_1$, and $V_1$, of B, R, and V neighbors, respectively.
Similarly, the numbers for the V agent are $D_2$, $R_2$, and $V_2$.
Then the value of $P$ before and after the switch
are:
\[
P_{initial}=2 B_1 + V_1 + B_2 + R_2; \quad P_{final}=2B_2 + V_2 +
B_1 + R_1.
\]
Thus,
\[
P_{final} - P_{initial} = B_2 + V_2 + R_1 - \left( B_1 + V_1 + R_2
\right).
\]
Taking into account that $B_1 + V_1 + R_1 = B_2 + V_2 + R_2 = 8$, we
arrive at
\[
P_{final} - P_{initial} = 2 \left( R_1 - R_2 \right) < 0.
\]
Similarly, if the switch between R and B, we have
\[
P_{final} - P_{initial} = 2 \left( R_2 - R_1 \right) + 2 \left( B_1 - B_2
\right) < 0.
\]

\medskip

\noindent 5. We define the {\it diameter of a cluster}, $L$, to be
the side length of the smallest square that covers the cluster. The
diameter of a cluster can be easily computed as the larger between
the number of rows that contain an agent belonging to the cluster
and the number of columns that contain an agent belonging to the
cluster. For configurations consisting of mostly compact clusters,
the maximum diameter, $L=\max(L_i)$, defines the {\it scale of
aggregation}.

\end{document}